\documentclass[print]{revtex4}
\usepackage{graphicx}

\begin{document}
\title{Dark States and Coherent Control of Spin States in Molecular Magnets}
\author{Jing-Min Hou$^{1}$\footnote{Electronic address: jmhou@eyou.com}, Li-Jun Tian$^2$, and Shuo Jin$^3$ }
\affiliation{$^1$Department of Physics, Southeast University, Nanjing, 210096, China\\
 $^2$Department of Physics,
Shanghai University, Shanghai, 200436, China\\
$^3$Department of Physics, School of Science, Beihang University,
Beijing, 100083, China}
\begin{abstract}
We propose a scheme to realize  coherent control of   spin states
of molecule magnet, Ni$_4$. We  introduce transverse magnetic
fields with special frequencies. When the frequencies of
transverse magnetic fields  match in some conditions, we obtain
dark states in Ni$_4$ molecules. Through adjusting the magnitude
of magnetic fields,
we can obtain any arbitrary superposition of the two ground spin states of Ni$_4$ molecules. \\
PACS number(s):42.50.Lc, 42.50.Hz, 75.50.Xx. 
\end{abstract}
\maketitle

 The dark state\cite{Arimondo} is a specific
coherent superposition of ground states that is immune to further
excitation due to destructive quantum interference. It has played
an important role in studies of a number of coherent phenomena in
physics such as electromagnetically induced transparency
\cite{Harris}, coherent population trapping \cite{Bergmann},
lasing without inversion \cite{Kocharovskaya}, slow propagation of
light in a medium\cite{Hau},  optical information
storage\cite{Fleischhauer},  and so on.

 Coherent population
transfer\cite{Shore} is another application of dark states. When
the dark states condition is met, via coherent population
transfer, one can produce excitation between states of the same
parity, for which single-photon transitions are forbidden for
electricdipole radiation, or between magnetic sublevels. The
coherent control of quantum states  is a central issue in the
emerging fields of spintronics and quantum information processing.
 Coherent population transfer is an effective method to realize
the coherent control of quantum states.

Recently, molecular magnets have received much attention because
of observed quantum features at the mesoscopic level and their
potential uses in magnetic storage and quantum computing.
Leuenberger and Loss have proposed a scheme to realize Grover's
algorithms in molecular magnets such as $\textrm{Fe}_8$ and
$\textrm{Mn}_{12}$\cite{Loss}. Hou et al. have suggested to
perform CNOT operation in dimer of molecular magnets,
[Mn$_4$]$_2$.\cite{Hou}

 In this paper, we propose a scheme to realize dark states in
 molecular magnet Ni$_4$\cite{Yang,Barco,Park}, so that we can obtain
an arbitrary superposition of the two spin states of Ni$_4$
molecule. Therefore  the coherent control of   spin states of
molecule magnets is realized. [Ni(hmp)(t-BuEtOH)Cl]$_4$, refereed
to as Ni$_4$, consists of four Ni$^{II}$(spin 1) magnetic ions and
oxygen atoms at alternating corners of a distorted cube, with
S$_4$ site symmetry. Ferromagnetic exchange interactions between
the Ni$^{II}$ ions lead to an $S=4$ ground state at low
temperature.

The corresponding Hamiltonian of  Ni$_4$ molecule is given by
\begin{equation}
H=H_0+H_1.
\end{equation}
Here $H_0$ is the Hamiltonian without the external field. It can
be written as
\begin{equation}
H_0=-D{\hat S}_z^2,
\end{equation}
where $D$ is the axial anisotropy constant and ${\hat S}_z$ is the
$z$ component of spin operator. $H_1$ is the Hamiltonian due to
the interactions with the applied magnetic field
\begin{equation}
H_1=-g\mu_B{\bf B}\cdot\hat{\bf S},
\end{equation}
where $\mu_B$ is the Bohr magneton, $g$ is the electronic
g-factor, $\hat{\bf S}$ is the spin operator, and $\bf B$ is the
applied magnetic field.

Each Ni$_4$ molecule can be modelled as a 'giant spin' of $S=4$
with Ising-like anisotropy. Every eigenstate of  Ni$_4$ can be
labelled by $|m\rangle$ with $m=4,3,\cdots, -4$. With the
projection operator $|m\rangle\langle m|$, the Hamiltonian without
the applied  external field can be written as
\begin{equation}
H_0=\sum_m\hbar\omega_m|m\rangle\langle m|,
\end{equation}
where  $\omega_m=-m^2D/\hbar$.

\begin{figure}[ht]
 \includegraphics[width=0.8\columnwidth]{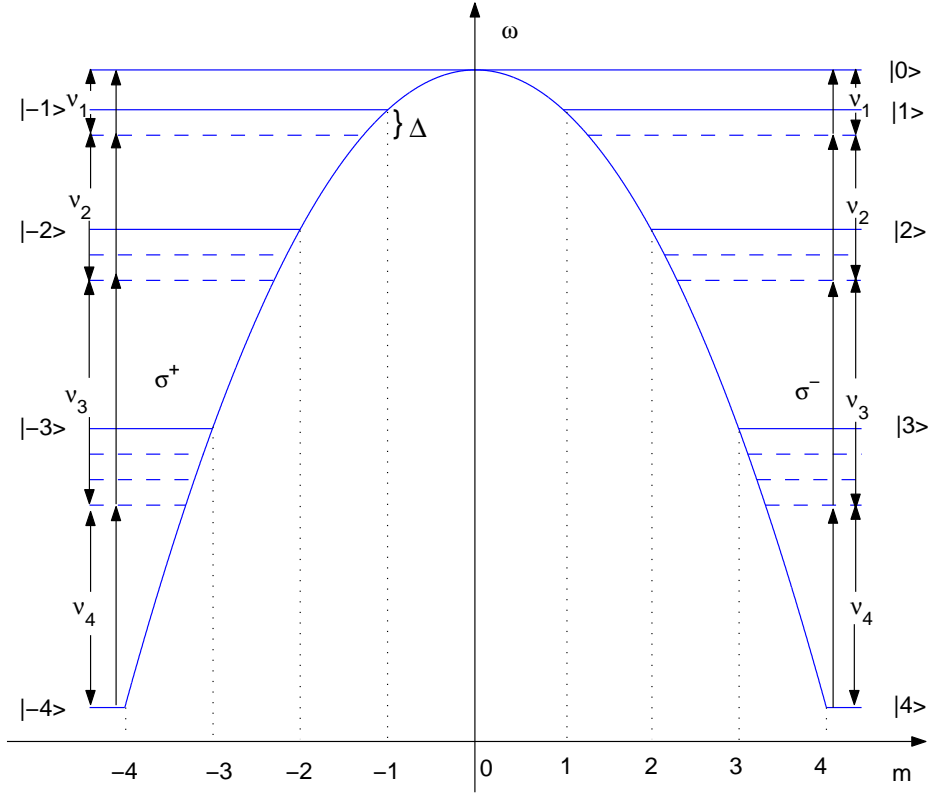}
 \caption{The  diagram of  energy levels of Ni$_4$ molecules.
 Here $\omega=E/\hbar$ represents the energy of spin states and $\nu_i$\ ($i=\pm1,\pm2,\pm3,\pm4$) are
 the frequencies of the applied transverse magnetic fields.
 $\sigma^+$ and $\sigma^-$ denote the left and right circularly polarized  photons respectively. $\Delta$ is  detuning. }
 \label{fig1}
\end{figure}

To realize coherent control of spin states of Ni$_4$, we introduce
time-dependent transverse magnetic fields $V_m(t)=B_m[\cos(\nu_m
t+\phi_m)\textbf{e}_x-\sin(\nu_m t+\phi_m)\textbf{e}_y]$ and
$V_{\bar{m}}(t)=B_{\bar{m}}[\cos(\nu_{\bar{m}} t+\phi_{\bar
m})\textbf{e}_x+\sin(\nu_{\bar m} t+\phi_{\bar m})\textbf{e}_y]$,
where $m=1,\cdots,4$ and $\bar{m}=-4,\cdots,-1$; $\phi_m$ and
$\phi_{\bar m}$ are phases of the transverse magnetic fields;
$\textbf{e}_x$ and $\textbf{e}_y$ are respectively the unit
vectors pointing along $x$ and $y$ axes. Thus we obtain the
Hamiltonian due to the interactions with the transverse magnetic
fields as
\begin{eqnarray}
\label{h1}
 H_1(t)&=& -g\mu_B\left\{\sum_{m=1}^4 B_m[\cos(\nu_m
t+\phi_m)\hat{S}_x-\sin(\nu_m
t+\phi_m)\hat{S}_y]+\sum_{\bar{m}=-4}^{-1}
B_{\bar{m}}[\cos(\nu_{\bar{m}} t+\phi_{\bar
m})\hat{S}_x+\sin(\nu_{\bar{m}}
t+\phi_{\bar m})\hat{S}_y]\right\}\nonumber\\
&=&-\frac{g\mu_B }{2}\left\{\sum_{m=1}^4B_m[e^{i(\nu_m
t+\phi_m)}\hat{S}_++e^{-i(\nu_m
t+\phi_m)}\hat{S}_-]+\sum_{\bar{m}=-4}^{-1}
B_{\bar{m}}[e^{-i(\nu_{\bar{m}} t+\phi_{\bar
m})}\hat{S}_++e^{i(\nu_{\bar{m}} t+\phi_{\bar
m})}\hat{S}_-]\right\}.
\end{eqnarray}
The transverse magnetic fields described by the first part in
Eq.(\ref{h1}) rotate clockwise and  produce left circularly
polarized $\sigma^-$ photons. Absorption (emission) of $\sigma^-$
photons gives rise to $\Delta m=-1(\Delta m=+1)$ transitions of
spin states. While the transverse magnetic fields described by the
second part in Eq.(\ref{h1}) rotate anticlockwise and produce
right circularly polarized $\sigma^+$ photons. Absorption
(emission) of $\sigma^+$ photons gives rise to $\Delta m=+1(\Delta
m=-1)$ transitions of spin states.

 There is a key point needed to mention. In order to
eliminate the excitation states in  the process, we choose the
special frequencies of the transverse magnetic fields with
appropriate detuning  from the gaps between spin states. Here we
choose  the  frequencies $\nu_i(i=-4,-3,-2,-1,1,2,3,4)$ of the
transverse magnetic fields
 far-detuning from energy gaps of
spin states, such as
$\omega_3-\omega_4-\nu_4=3\Delta,\omega_2-\omega_3-\nu_3=-\Delta,
\omega_1-\omega_2-\nu_2=-\Delta,\omega_0-\omega_1-\nu_1=-\Delta,
\omega_0-\omega_{-1}-\nu_{-1}=-\Delta,\omega_{-1}-\omega_{-2}-\nu_{-3}=-\Delta,
\omega_{-2}-\omega_{-3}-\nu_{-3}=-\Delta,\omega_{-3}-\omega_{-4}-\nu_{-4}=3\Delta$
and $\nu_i=\nu_{-i}$\ ($i=1,2,3,4$), where $\Delta$ is a finite
detuning.  The detail of the choosing of the frequencies $\nu_i$
is shown in detail in Figure \ref{fig1}. In Reference \cite{Park},
measured EPR spectra on the Ni$_4$ molecules reveal that values of
$D$ are in the range of $0.72-1.03$ K. So, the energy gaps between
different spin states are  $\sim 10^{-23}$J. We estimate that the
frequencies of the transverse magnetic fields are in the range of
$100\sim$700GHz.

Next we calculate the quantum amplitudes for the transitions
induced by the time-dependent transverse magnetic  fields  by
evaluating the $S$-matrix perturbatively\cite{Loss}. The
$S$-matrix can be expanded in the perturbation series in power of
the interaction Hamiltonian $H_1^I(t)=e^{iH_0
t/\hbar}H_1(t)e^{-iH_0 t/\hbar}$ as $S=\sum_{n=0}^\infty S^{(n)}$.
The $n$th-order term of the perturbation series of $S$-matrix is
expressed by
\begin{equation}
S^{(n)}=\left(-\frac{i}{\hbar}\right)^n\int_{-\infty}^{+\infty}dt_1\int_{-\infty}^{t_1}dt_2\cdots\int_{-\infty}^{t_{n-1}}
dt_nH_1^I(t_1)H_1^I(t_2)\cdots H_1^I(t_n).
\end{equation}
In the $H_0$ representation, the $S$-matrix is expressed by
$S_{\beta\alpha}=\sum_{n=0}^\infty S_{\beta\alpha}^{(n)}$, which
means the transition amplitude from  state $\alpha$ to state
$\beta$.  After a straightforward calculation, we find that the
perturbation terms below the fourth order vanish. We also obtain
the nonzero fourth order perturbation terms as follow,
\begin{eqnarray}
S_{0,4}^{(4)}&=&\frac{ie^{-i(\phi_1+\phi_2+\phi_3+\phi_4)}\prod_{i=1}^4\Omega_i}{96\Delta^3}\int_{-\infty}^\infty
dt,\\
S_{0,-4}^{(4)}&=&\frac{i
e^{-i(\phi_{-1}+\phi_{-2}+\phi_{-3}+\phi_{-4})}
\prod_{\bar{m}=-4}^{-1}\Omega_{\bar{m}}}{96\Delta^3}\int_{-\infty}^\infty
dt.
\end{eqnarray}
where $\Omega_i=g\mu_B B_i/\hbar\ (i=-4,-3,-2,-1,1,2,3,4)$.  The
higher order perturbation terms are  negligible compared with
$S_{0,4}^{(4)}$ and $S_{0,-4}^{(4)}$, so we shall not consider the
contribution of the higher order perturbation terms in our
approximation.

 We can interpret the above perturbation series of
$S$-matrix as the below explanation. Because we choose the
appropriate frequencies of the applied transverse magnetic fields
with detuning, the transition rates between some spin states are
so small that they can be neglected. Only the transitions between
spin states $|4\rangle$ and $|0\rangle$ and between spin states
$|-4\rangle$ and $|0\rangle$ are obvious since the  frequencies of
the applied transverse magnetic fields match the gaps of these
spin states in four-photon process as shown in Figure \ref{fig1}.
These transitions are four order processes, i.e. when the
transitions happen, four photons are absorbed or emitted. If the
molecules only occupy the spin states $|4\rangle$ and $|-4\rangle$
initially, so, after the applied transverse magnetic field are
added, only the populations of the spin states $|4\rangle,
|-4\rangle$ and $|0\rangle$ are nonzero. Therefore,  Ni$_4$
molecule can be considered as a three-level atom. Suppose that we
have an effective Hamiltonian $H^{I}_{eff}$ in the interaction
picture about the three-level atom model and that the $S$-matrix
of it is equal to that of $H_1^I$. Thus, with this approximation,
we obtain the effective Hamiltonian
  in the interaction picture as,
\begin{eqnarray}
H^{I}_{eff}=-\frac{\hbar}{2}[\Omega_a(e^{-i\phi_a}|0\rangle\langle4|+e^{i\phi_a}|4\rangle\langle0|)
+\Omega_b(e^{-i\phi_b}|0\rangle\langle-4|+e^{i\phi_b}|-4\rangle\langle0|)]
\label{EH}
\end{eqnarray}
where $\phi_a=\sum_{m=1}^4\phi_m$ and
$\phi_b=\sum_{\bar{m}=-4}^{-1}\phi_{\bar{m}}$; $\Omega_a$ and
$\Omega_b$ are the effective Rabi frequencies written as
\begin{eqnarray}&&\Omega_a=\frac{\prod_{i=1}^4\Omega_i}{48\Delta^3},\label{Rabi1}\\&&\Omega_b=\frac{
\prod_{\bar{m}=-4}^{-1}\Omega_{\bar{m}}}{48\Delta^3}.\label{Rabi2}\end{eqnarray}

 We consider the Ni$_4$ molecule as a three-level atom described
by the effective Hamiltonian (\ref{EH}). For convenience, here we
set $\phi_b=0$ and $\phi_a-\phi_b=\pm\pi$. It is straightforward
to verify that the following combinations of spin states
$|4\rangle, |-4\rangle$, and $|0\rangle$ are eigenstates of this
Hamiltonian \cite{Shore,Scully}:
\begin{eqnarray}
&&|\Psi_+\rangle=\frac{1}{\sqrt{2}}(|0\rangle-\sin\theta|4\rangle +\cos\theta|-4\rangle),\\
&&|\Psi_0\rangle=\cos\theta|4\rangle+\sin\theta|-4\rangle,\label{DS}\\
&&|\Psi_-\rangle=\frac{1}{\sqrt{2}}(|0\rangle+\sin\theta|4\rangle
-\cos\theta|-4\rangle),
\end{eqnarray}
where $\cos\theta=\Omega_b/\sqrt{\Omega_a^2+\Omega_b^2},\
\sin\theta=\Omega_a/\sqrt{\Omega_a^2+\Omega_b^2}$; $\theta$ is
mixing angle. The corresponding eigenvalues are $E_\pm=\mp
\frac{\hbar}{2}\sqrt{\Omega_a^2+\Omega_b^2}$ and $E_0=0$.
  Here, the wave function $|\Psi_0\rangle$
 is the so-called dark state.  From Eq.(\ref{DS}), we
can see that in the dark state the populations of the spin states
$|4\rangle$ and $|-4\rangle$ do not vary with time and no
excitation to the spin state $|0\rangle$ happen. The population is
trapped in the lower spin states. Although the transverse magnetic
fields exist, there is no absorbtion due to the destructive
quantum interference between the two transitions.

\begin{figure}[ht]
 \includegraphics[width=0.8\columnwidth]{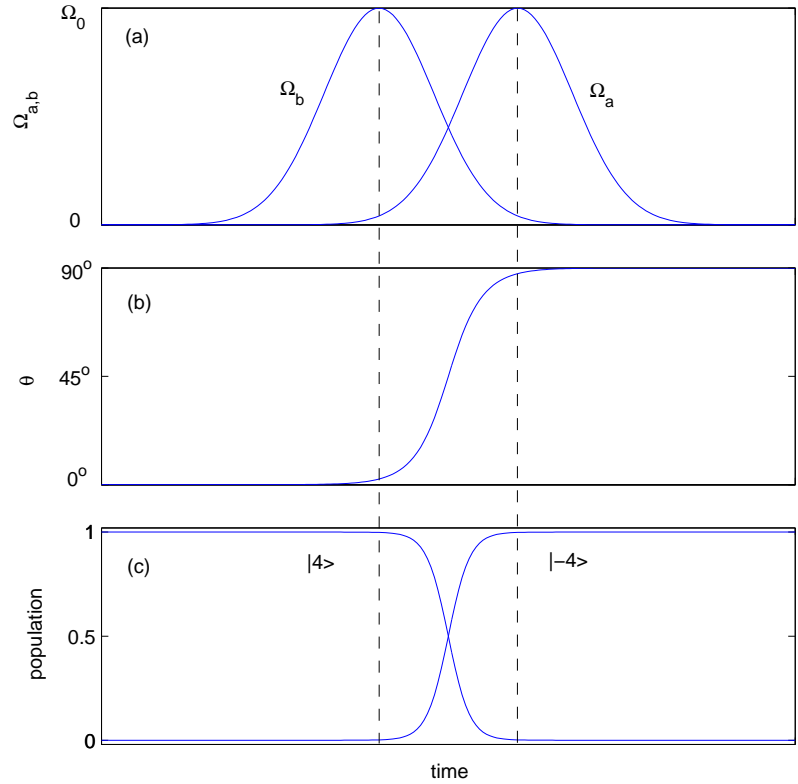}
 \caption{The  diagram of  coherent control of spin states
 $|4\rangle\rightarrow|-4\rangle$. (a)the Rabi frequencies; (b) the mixing angle; (c) the population of spin states }
 \label{fig2}
\end{figure}

When the Rabi frequency $\Omega_b$ is nonzero and $\Omega_a=0$,
i.e. the mixing angle $\theta=0$, the spin state $|4\rangle$ is
right the dark state $|\Psi_0\rangle$. In this condition, the gaps
between the dark states and the eigenstates $|\Psi_+\rangle$ and
$|\Psi_-\rangle$ are $\pm\frac{\hbar}{2}\Omega_b$, which are
nonzero. We choose the spin state $|4\rangle$ as the initial
state. At first we adiabatically increase $\Omega_b$ from zero to
a finite value and keep $\Omega_a$ zero. Then we
 change the Rabi frequencies $\Omega_a$ and $\Omega_b$ as we want via varying the amplitudes of the
transverse magnetic fields, so the mixing angle $\theta$ will vary
continually with time. Therefore the state can be written as
\begin{equation}
|\Psi_0(t)\rangle=\frac{\Omega_b(t)|4\rangle+\Omega_a(t)|-4\rangle}{\sqrt{\Omega_a(t)+\Omega_b(t)}}
\end{equation}
where $\Omega_a(t)$ and $\Omega_b(t)$ are time-dependent functions
that vary adiabatically with time.  We note that  any arbitrary
superposition of the two ground spin states $|4\rangle$ and
$|-4\rangle$ can be prepared by appropriately tailoring the shapes
of pulse of the transverse magnetic fields. That is, we can
coherently control the spin states of molecular magnets by
adiabatically changing the Rabi frequencies.

\begin{figure}[ht]
 \includegraphics[width=0.7\columnwidth]{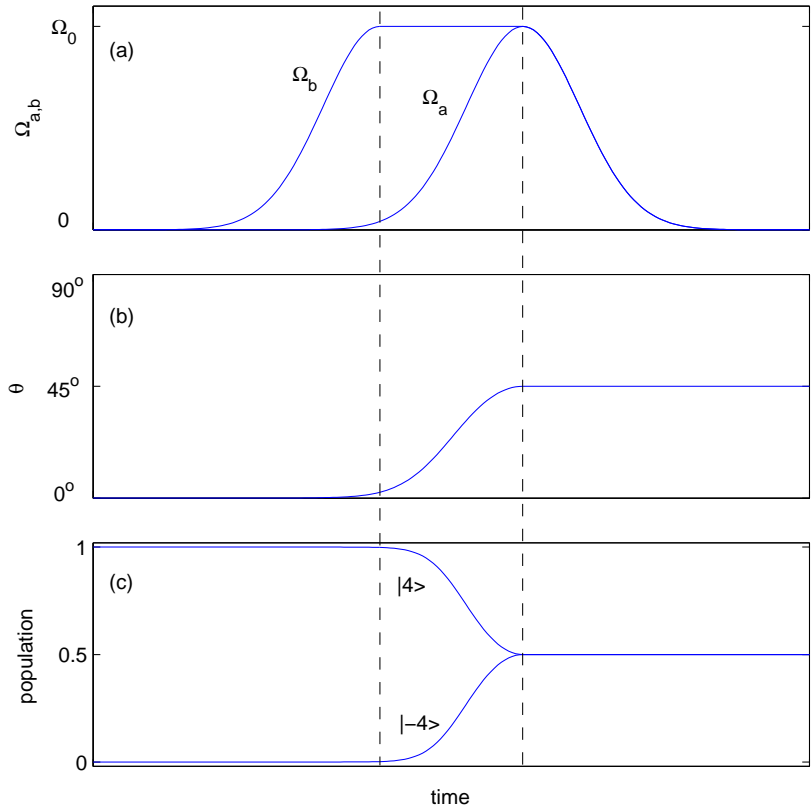}
 \caption{The  diagram of  coherent control of spin states
 $|4\rangle\rightarrow\frac{1}{\sqrt{2}}(|4\rangle+|-4\rangle)$. (a)the Rabi frequencies; (b) the mixing angle; (c) the population of spin states}
 \label{fig3}
\end{figure}

In order to clearly explain the process of coherent control of
spin states, we take  two special examples that are showed in
 Figures \ref{fig2} and \ref{fig3} respectively.
In the case showed in Figure \ref{fig2}, the initial state is
$|4\rangle$. In the beginning, the Rabi frequency $\Omega_b$
adiabatically increases from zero and $\Omega_a$ keeps zero. It
easy to verify that the initial state is right the dark state at
the moment. Because only the population of the spin state
$|4\rangle$ is nonzero, the mixing angle $\theta$ is zero. After
$\Omega_b$ increases to some value, we can decrease $\Omega_b$ and
increase $\Omega_a$ from zero adiabatically . When $\Omega_b$
vanishes, $\Omega_a$ arrives to a finite value. Simultaneously,
the mixing angle $\theta$ changes from 0 to $\pi/2$, so the
coherent population transfer is realized from $|4\rangle$ to
$|-4\rangle$, i.e., one can coherently control the evolution of
the spin state from $|4\rangle$ to $|-4\rangle$ by this scheme. In
Figure \ref{fig3}, we show another case that the coherent and
adiabatic evolution of  the spin state from $|4\rangle$  to
$\frac{1}{\sqrt{2}}(|4\rangle+|-4\rangle)$ is realized. In this
case,  firstly, we increase $\Omega_b$ to a finite value and
$\Omega_a$ is zero as the previous case. Next, we keep $\Omega_b$
invariant and adiabatically increase $\Omega_a$ to the same value
from zero. Finally, we decrease $\Omega_a$ and $\Omega_b$ to zero
simultaneously. The last two steps are different from that of the
previous case and the mixing angle varies from 0 to $\pi/4$. In
the end, the populations of the spin states $|4\rangle$ and
$|-4\rangle$ are both $1/2$. From Figure \ref{fig3} we can see
 that the coherent control of spin states
from $|4\rangle$ to $\frac{1}{\sqrt{2}}(|4\rangle+|-4\rangle)$ is
realized by our scheme.

\begin{figure}[ht]
 \includegraphics[width=0.7\columnwidth]{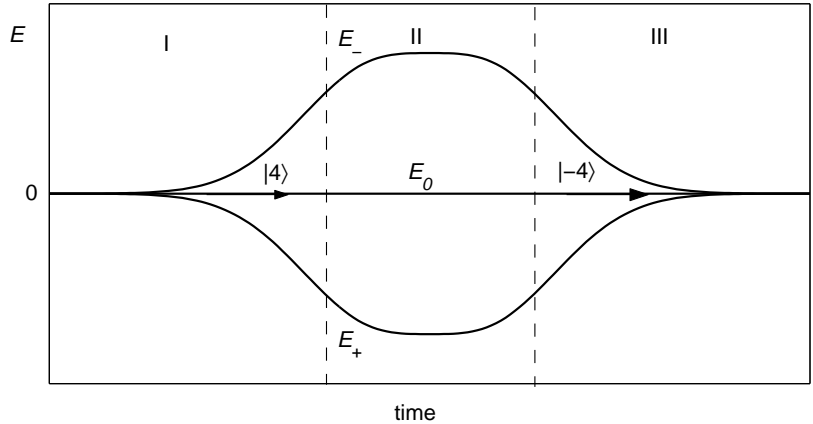}
 \caption{The variation of  eigenvalues of the three eigenstates of the Hamiltonian (\ref{EH})
 with time
 in the condition presented in Figure \ref{fig2}. }
 \label{fig4}
\end{figure}

Until now, we have not yet discussed the adiabatic condition for
evolution of the dark state in this paper. In order to interpret
adiabatic evolution of the dark state, we take the case shown in
Figure \ref{fig2} as example. In Figure \ref{fig4}, we draw the
variation of eigenvalues of the three eigenstates with time in the
condition shown in Figure \ref{fig2}. At the initial time, all the
transverse magnetic fields are not added, $|4\rangle$ is just the
dark state $|\Psi_0\rangle$ and the three eigenstates are
degenerate. In interval I, $\Omega_b$ is added and $\Omega_a$
keeps zero, the energy gaps between the dark state and the other
eigenstates appear but the eigenvalue of the dark state
$|\Psi_0\rangle$ remains unchanged. During interval II, both
$\Omega_a$ and $\Omega_b$ are nonzero, the dark state has
$|4\rangle$ and $|-4\rangle$ components and the energy splitting
of eigenvalues $E_\pm$ and $E_0$ is largest. In interval III, the
Rabi frequency $\Omega_b$ is zero and $\Omega_a$ gradually
decreases to zero, the energy splitting of eigenenergies gradually
disappears. From above we can see that there always exist gaps
between eigenvalues $E_\pm$ and $E_0$ in our operation time. In
other cases besides the example, the processes are similar. Thus,
if we choose the spin state $|4\rangle$, which is right the dark
state in interval I,  as the initial state, it is possible to
evolve adiabatically the dark state to our requested quantum state
by adding special transverse magnetic field pulses.

 In order to realize adiabatic evolution of the dark state, it is
 necessary  that the amplitudes of transverse magnetic fields are large enough.
  If the
coupling is insufficient, i.e. the Rabi frequencies are too small,
the practical quantum state  can not follow the evolution of the
dark state $|\Psi_0\rangle$ and loss of population due to
nonadiabatic transfer to the states $|\Psi_+\rangle$and
$|\Psi_-\rangle$ may occur. The Hamiltonian matrix element for
nonadiabatic coupling between the state $|\Psi_0\rangle$ and the
states $|\Psi_+\rangle$ or $|\Psi_-\rangle$ is given by $\langle
\Psi_\pm|\dot\Psi_0\rangle$ \cite{Messiah}. Thus, the adiabatic
condition is \cite{Shore}
\begin{equation}
\left|\langle \Psi_\pm|\dot\Psi_0\rangle\right|\ll
\frac{\left|E_\pm-E_0\right|}{\hbar}, \label{adiabatic}
\end{equation}
When the transverse magnetic field pulses have a smooth shape, a
convenient adiabaticity criterion may be derived from Eq.
(\ref{adiabatic}) by taking a time average of the left-hand side,
$\langle \Psi_\pm|\dot\Psi_0\rangle_{av}=\pi/2\Delta\tau$, where
$\Delta\tau$ is the period during which the pulses overlap. This
average value should not exceed the right-side,
$\left|E_\pm-E_0\right|/\hbar=\sqrt{\Omega_a^2+\Omega_b^2}/2$. The
adiabatic condition is \cite{Shore}
\begin{equation}
\Omega_{av}\Delta\tau>10\label{avrabi}
\end{equation}
where $\Omega_{av}$ is the average Rabi frequency of $\Omega_a$
and $\Omega_b$, and the number 10 is obtained from experience and
numerical simulation studies.

The experimental work \cite{Barco} shows that, in Ni$_4$
molecules, the longitudinal relaxation time is long, and its order
is $\sim s$. Here we assume that $\Delta\tau$ has the same order
with the relaxation time. From this relaxation time and
Eq.(\ref{avrabi}), we obtain the restriction for the average Rabi
frequencies is $\Omega_{av}>10s^{-1}$. In our scheme, it is proper
that we choose the detuning $\Delta$ as 1GHz. Thus, from
Eq.(\ref{Rabi1}) or (\ref{Rabi2}), we know that $
\bar{\Omega}_i>2.6\times 10^{7}s^{-1} $, where
$\bar{\Omega}_i=g\mu_B \bar{B}_i/\hbar$ is the average value of
$\Omega_i= g\mu_B {B}_i/\hbar$. Here we introduce the average
transverse magnetic field $\bar{B}_i$. Reference \cite{Yang} shows
$g\sim 2.2$. Finally we obtain the amplitude of the transverse
magnetic fields required in our scheme as $ \bar{B}_i>8.4\times
10^{-4}T $. In our scheme, the precise matching between the sum of
frequencies of  the left circularly polarized transverse magnetic
fields $\sum_{i=1}^4 \nu_i$ and the sum of frequencies of  the
right circularly polarized transverse magnetic fields
$\sum_{\bar{i}=-4}^{-1}\nu_{\bar{i}}$ is required, but a finite
detuning between $\sum_{i=1}^4 \nu_i$ and the gap of the spin
states $|0\rangle$ and $|4\rangle$, $\omega_0-\omega_4$, or
between $\sum_{\bar{i}=-4}^{-1} \nu_{\bar{i}}$ and the gap of the
spin states $|0\rangle$ and $|-4\rangle$, $\omega_0-\omega_{-4}$,
does not prevent the dark resonance.  From the above estimation,
we can see that the conditions required in our scheme are
accessible experimentally.

In conclusion, using the method suggested by us, one can prepare
any arbitrary superposition of the spin states $|4\rangle$ and
$|-4\rangle$ from the ground spin state $|4\rangle$ or
$|-4\rangle$. In molecular magnets, there is a potential barrier
between the spin states with positive magnetic numbers and with
negative ones, as is shown in Figure \ref{fig1}. In the absence of
longitudinal magnetic field, each molecule has a double, Kramers
degeneracy in its ground states. In Ni$_4$ molecules, they are
$|4\rangle$ and $|-4\rangle$. The transition from one ground state
to the other is possible only if the molecule jumps over the
potential barrier or tunnels through it. Moreover, it is difficult
to coherently control the transition caused  by jumping over  or
tunnelling through the potential barrier and these schemes are not
spin-state-selective. However, our scheme is spin-state-selective
and the process is coherent. Therefore, Through adjustment of the
magnitudes of the transverse magnetic fields, any arbitrary
superposition of the two spin states of Ni$_4$ molecule can be
obtained, i.e., the coherent control of   spin states in molecular
magnet, Ni$_4$, is realized.

\begin{acknowledgments} This project  is supported  by NSF of China Grant Nos.10547107, 10571091, and Doctoral
Starting Fund from Southeast University No.9207022244. We thank
referees for their helpful advices to improve the paper.
\end{acknowledgments}

\end{document}